# Dynamics of Pure Spin Current in High-frequency Quantum Regime

Shuichi Iwakiri,[1] Yasuhiro Niimi,[1,2] and Kensuke Kobayashi[1,2]

[1]*Department of Physics, Graduate School of Science, Osaka University, Osaka 560-0043, Japan*

[2]*Center for Spintronics Research Network (CSRN), Graduate School of Engineering Science, Osaka University, Osaka 560-8531, Japan.*

## Abstract

Pure spin current is a powerful tool for manipulating spintronic devices, and its dynamical behavior is an important issue. By using mesoscopic transport theory for electron tunneling induced by spin accumulation, we investigate the dynamics of the spin current in the high-frequency quantum regime, where the frequency is much larger than temperature and bias voltage. Besides the thermal noise, frequency-dependent finite noise emerges, signaling the spin current across the tunneling barrier. We also find that the autocorrelation of the spin current exhibits sinusoidal oscillation in time as a consequence of the Pauli exclusion principle even without any net charge current.

Spin current, a flow of spin angular momentum, is one of the central concepts in spintronics [1-3]. It plays an important role not only as a tool for controlling various spintronic devices but also as a probe for spin dynamics in solids [1-9]. Especially, pure spin current, a flow of spin without net charge current, is of great interest because it is supposed to be dissipationless [10-12]. Besides various studies to utilize it, there are several attempts to reveal its fundamental nature [13,14,15]. Nevertheless, there remains much to be done to understand the spin current itself and new measurement probes are required. As spin is a quantum object, it is of significance to address quantum nature of the spin current by investigating its dynamics [15].

Shot noise, or non-equilibrium current fluctuation, can be a unique tool to tackle this problem. Conventionally, it appears when electrons tunnel through a potential barrier such that the current noise power spectral density $S_I$ is expressed as the Schottky formula $S_I = 2e|I|$ in the zero-frequency and zero-temperature limit ($I$: the current and $e$: electric charge) [16]. Interestingly, shot noise is not trivial for spin current [17,18]. Consider a pure spin current, where the current with spin-up electrons ($I_\uparrow$) and that with spin-down ones ($I_\downarrow$) move in the opposite direction (say, $I_\uparrow = -I_\downarrow$). While there is no net charge current, the fluctuation of each current adds up according to the Schottky formula, yielding a finite shot noise of $S_I = 2e(|I_\uparrow| + |I_\downarrow|)$. Recently, the relevance of this concept of "spin shot noise" was experimentally demonstrated in the (Ga(Mn)As-based all-semiconductor lateral spin-valve device [18]. On the other hand, the above treatment is made for the zero-frequency limit and the shot noise at the finite frequency would enable us to obtain new information on transport dynamics.

Recently, the shot noise in this regime was experimentally addressed [19,20,21]. Thibault *et al.* [19] clarified an important aspect of the current fluctuation in the charge transport in a tunnel barrier. They measured noise spectra in the high-frequency quantum regime, namely, the frequency region much higher than both temperature and bias voltage, and derived the autocorrelation of the charge current. They observed the oscillation of autocorrelation in time as a direct consequence of the Pauli exclusion principle in the current. This is a clear manifestation of the quantum nature of the charge current. However, such an attempt has been lacking for the spin current.

In this Letter, expanding the concept of the spin shot noise to the finite frequency region, we show that the Pauli exclusion principle is also relevant for the spin current even when there is no net charge current. We evaluate the noise spectrum and autocorrelation of the spin current induced by the spin accumulation at the tunnel barrier. We also discuss the experimental feasibility of our results.

Figure 1(a) shows the setup we consider here, which is similar to the conventional lateral-spin valve system [23,24]. The device consists of one ferromagnetic lead (FM) that is magnetized along the lead, and two nonmagnetic leads attached to it; middle lead (NM1) and right lead (NM2). There is a tunnel barrier between NM1 and NM2. Note that we assume NM2 to be nonmagnetic just for simplicity, while the following result can be easily generalized for the ferromagnetic case by taking the spin polarization into account phenomenologically.

By injecting a spin-polarized current from FM to NM1, spin accumulation is created inside NM1 along the *x*-axis (Fig.1 (a)); The chemical potentials of the spin-up and spin-down electrons become *x*-dependent, generating a spin current as shown in Fig. 1(b). A part of this spin current flows down along the *z*-axis into NM2 through the tunnel barrier. The energy diagram in the vicinity of the barrier is presented in Fig. 1(c), where the chemical potentials of the spin-up and spin-down electrons steeply change at the barrier. The resulting potential difference at the barrier corresponds to the spin accumulation ($\delta\mu$). Here, we neglect the effect of the spin diffusion along the *z*-axis across the barrier, as the signature of such an effect was never detected in the experiment of Ref.[18]. The spin diffusion does not play an important role as long as we consider devices similar to those in Ref.[18]. We apply $V$ as the voltage difference between NM1 and NM2 (Figure 1(c) shows the case of $V = 0$). We consider the noise generated at this barrier. Note that, as it occurs locally, the noise is irrelevant to the spin diffusion process in NM2. The spin flip during the tunneling can be neglected, which was validated by the recent experiment [18].

The calculation is performed using the mesoscopic transport theory based on the Landauer-Büttiker formalism [24]. The tunnel barrier is treated as a one-dimensional single channel scatterer with an energy-independent transmission probability, $\tau$. Note that it is straightforward to extend our analysis to the multi-channel case. We define a current operator $\hat{I}_{\gamma,\sigma}$ using the second quantization. Here, $\gamma$ and $\sigma$ denote leads (L:NM1 or R:NM2) and spin (↑ or ↓), respectively. By taking the quantum statistical average, the mean current is given in the well-known formula as follows:

$$\langle \hat{I}_{\gamma,\sigma} \rangle = \frac{e}{h}\tau \int_{-\infty}^{\infty} dE [f_{\gamma,\sigma}(E,\mu_{\gamma,\sigma},T) - f_{\gamma',\sigma}(E,\mu_{\gamma',\sigma},T)], \quad (1)$$

where $h$ is the Planck constant, $E$ is the electron energy, and $f_{\gamma,\sigma}(E)$ is the Fermi distribution function for electrons with chemical potential $\mu_{\gamma,\sigma}$ and temperature $T$.

The spin current operator is defined as $\hat{I}_S = \hat{I}_\uparrow - \hat{I}_\downarrow$ [17]. Here, the spin-up and spin-down channels are independent of each other as we can neglect the spin flip. By integrating the Fermi distribution function, we obtain $\langle \hat{I}_{\gamma,\sigma} \rangle = \frac{e}{h}\tau(\mu_{\gamma,\sigma} - \mu_{\gamma',\sigma})(1 - \delta_{\gamma,\gamma'})$. Substituting the chemical potential of each lead $\mu_{L\uparrow/\downarrow} = \mu_0 + \frac{eV}{2} \pm \frac{\delta\mu}{2}$ and $\mu_{R\uparrow/\downarrow} = \mu_0 - \frac{eV}{2}$, we obtain $\langle \hat{I}_C \rangle = \frac{2e^2}{h}\tau V$ for the charge current, and $\langle \hat{I}_S \rangle = \frac{e}{h}\tau(eV +$ $\delta\mu$) for the spin current. This is consistent with the previous results [18].

Now we discuss the noise spectrum. By defining the current noise operator $\delta\hat{I}_{\gamma,\sigma} = \hat{I}_{\gamma,\sigma} - \langle\hat{I}_{\gamma,\sigma}\rangle$, the noise spectrum is expressed as follows:

$$S_{\gamma,\sigma}(\nu) = \int_{-\infty}^{\infty} \langle\delta\hat{I}_{\gamma,\sigma}(t)\delta\hat{I}_{\gamma,\sigma}(t+\tau)\rangle e^{2\pi i\nu\tau} d\tau,$$

where $t$ is time and $\nu$ is frequency defined from $-\infty$ to $+\infty$. Following the work by Meair *et al.* [17], in the framework of the Landauer-Büttiker formalism, the spin-dependent transmission channels can be treated as if they form a parallel circuit. Thus the total noise spectrum of the spin current $S_{total} = \sum_{\gamma,\sigma} S_{\gamma,\sigma}(\nu)$ is analytically given as follows:

$$S_{total}(\nu,T,V,\delta\mu) = \frac{e^2}{h}\frac{4\tau^2 h\nu}{1-e^{-\frac{h\nu}{k_B T}}} + \frac{e^2}{h}\frac{\tau(1-\tau)\left(eV+\frac{\delta\mu}{2}+h\nu\right)}{1-e^{-\frac{1}{k_B T}\left(h\nu+eV+\frac{\delta\mu}{2}\right)}} + \frac{e^2}{h}\frac{\tau(1-\tau)\left(-eV-\frac{\delta\mu}{2}+h\nu\right)}{1-e^{-\frac{1}{k_B T}\left(h\nu-eV-\frac{\delta\mu}{2}\right)}}$$
$$+ \frac{e^2}{h}\frac{\tau(1-\tau)\left(eV-\frac{\delta\mu}{2}+h\nu\right)}{1-e^{-\frac{1}{k_B T}\left(h\nu+eV-\frac{\delta\mu}{2}\right)}} + \frac{e^2}{h}\frac{\tau(1-\tau)\left(-eV+\frac{\delta\mu}{2}+h\nu\right)}{1-e^{-\frac{1}{k_B T}\left(h\nu-eV+\frac{\delta\mu}{2}\right)}}. \qquad (2)$$

By taking the zero frequency and zero temperature limit, we obtain the expression consistent with that was given in Ref.[18]. The quantum nature of the current appears in the finite-frequency component of the calculated noise [24]. As it is generated via the process where there is a finite energy difference between the initial and the final states, the system absorbs/emits a photon to conserve the energy. Actually, Eq.(2) is understood in terms of one dimensional emission ($\nu < 0$) and absorption ($\nu > 0$) spectrum of photon with an energy of $h\nu \pm eV + \frac{\delta\mu}{2}$ for the spin-up channel and $h\nu \pm eV - \frac{\delta\mu}{2}$ for the spin-down channel. Thus the quantum nature is naturally introduced when we consider the finite frequency noise. When the emission and absorption processes occur with the same probability, the noise spectrum can be symmetrized with regard to the positive and negative frequency: $S_{\text{sym}}(\nu, T, V, \delta\mu) = S(\nu, T, V, \delta\mu) + S(-\nu, T, V, \delta\mu)$ with $\nu = [0, \infty)$.

In the rest of this paper, we focus only on the zero bias regime ($V = 0$, see Fig. 1(c)), where there is no net charge current across the barrier. For simplicity, we redefine $S_{\text{sym}}(\nu, T, \delta\mu) \equiv S_{\text{sym}}(\nu, T, V = 0, \delta\mu)$. Two remarks are made for the zero-frequency limit. First, $S_{\text{sym}}(\nu = 0, T, \delta\mu = 0) \equiv S_0$ gives the classical thermal (Johnson-Nyquist) noise [25]. Second, $S_{\text{sym}}(\nu = 0, T, \delta\mu)$ reproduces the previous result [18]. Now, for the finite frequency, in Fig.2, we show $S_{\text{sym}}(\nu, T, \delta\mu)/S_0$ as a function of the normalized frequency ($h\nu/k_B T$) for the cases of $\delta\mu = 0$ (no spin accumulation) and $\delta\mu/k_B T = 1$, 3, and 5 (finite spin accumulation). While $S_{\text{sym}}(\nu, T, \delta\mu = 0)$ again equals to the well-known thermal noise spectra, we found the increase of the noise when there is finite spin accumulation. This indicates that the shot noise is generated by the spin current even without any net charge current.

What does the increase of the noise mean? To understand this, we investigate the dynamics of the system in real time. Applying Wiener-Khinchin theorem to Eq. (2), we derive the autocorrelation of the spin current. Before this treatment, following Ref. [19], we need to redefine the current noise spectral density as $S'_{\text{sym}}(\nu, T, \delta\mu) \equiv S_{\text{sym}}(\nu, T, \delta\mu) - S_{\text{sym}}(\nu, T = 0, \delta\mu)$, because $S_{\text{sym}}$ diverges such that $S_{\text{sym}} \to 4e^2\tau\nu$ for $\nu \to \infty$. This subtraction means that we ignore the contribution of the vacuum fluctuation in the noise spectrum (see the dotted line in Fig.(2)), and focus ourselves only on the noise generated by the electron tunneling.

We found that the autocorrelation of the spin current is given as follows:

$$C(t, \delta\mu, T) = 4\tau\left((1-\tau)\cos\left(\frac{\delta\mu t}{\hbar}\right) + \tau\right)\left(\frac{1}{2}\left(\frac{\hbar}{k_B T}\right)^2 - 2\left(\frac{\pi}{e^{\frac{k_B T}{2\hbar}t} - e^{-\frac{k_B T}{2\hbar}t}}\right)^2\right). \quad (3)$$

We plot Eq. (3) as a function of time in units of $h/\delta\mu$ in Fig. 3. In the absence of the spin accumulation ($\delta\mu = 0$), the thermal noise is the only origin of the noise, making the autocorrelation $C(t, \delta\mu = 0, T)$ monotonously decrease according to time. This means that the quantum coherence of electron decays due to the thermal agitation with a characteristic time scale of $h/k_B T$. In the presence of the spin accumulation ($\delta\mu \neq 0$), the autocorrelation oscillates with the envelope $C(t, \delta\mu = 0, T)$. Thibault *et al.* observed a similar oscillation when a charge current flows across the voltage-biased barrier, which directly indicates that electron can tunnel only one by one due to the Pauli exclusion principle [19]. In the same way, the present oscillation suggests that this principle also affects transport dynamics of the spin current even when there is no net charge current.

Moreover, we found an interesting temperature-independent relation:

$$\frac{C(t,\delta\mu,T)}{C(t,0,T)} = \tau + (1-\tau)\cos\left(\frac{\delta\mu}{\hbar}t\right). \quad (4)$$

This shows that the oscillation with a frequency $\delta\mu/h$ is always present regardless of temperature.

The time dependence of the autocorrelation clearly shows the quantum nature of the spin current. With the same analogy as discussed in Ref. [19], the oscillation of the autocorrelation can be understood as follows. The tunneling of the spin current occurs as spin-up and spin-down electrons sequentially come into the barrier. Due to the Pauli's exclusion principle, only one spin-up and one spin-down electron can tunnel in a certain time. Here, spin-up and spin-down electrons have the same chemical potential difference with the same absolute value but with the opposite signs. In the quantum regime, because the energy before the tunneling and that after the tunneling are different, it takes a finite time to resolve the electron's wave function of the two states, which is at least $h/\delta\mu$ reflecting Heisenberg's uncertainty principle.

Thus the autocorrelation oscillates with the period of $h/\delta\mu$, indicating that the Pauli's exclusion principle acts on the spin current. Such quantum nature can be destroyed due to the decoherence processes. In this case, because of the thermal agitation, the coherence vanishes as a function of time, and thus the oscillation of the autocorrelation decay as shown in Fig. 3. Here, $h/k_B T$ has a dimension of time, which corresponds to the coherence time.

Then we discuss the feasibility of the experiment. The measurement frequency must be higher than $\delta\mu/h$ to observe more than one period of the oscillation. At the same time, it must be higher than $k_B T/h$ to inhibit the thermal decay of electron coherence. For example, when $T = 30$ mK and $\delta\mu = 5$ μeV, the noise spectra need to be measured up to the frequency higher than 1 GHz. This can be realized using dilution refrigerator (its temperature range is typically from a few ten mK to room temperature) with high frequency noise measurement setup. For example, measurement sensitivity can reach up to $10^{-27}$ A²/Hz while the shot noise of the sample with 50 ohm and 5 μeV bias is of order of $10^{-26}$ A²/Hz according to the Schottky formula.

to $10^{-27}$ $A^2/Hz$ while the shot noise of the sample with 50 ohm and 5 μeV bias is of order of $10^{-26}$ $A^2/Hz$ according to the Schottky formula.

      Finally, we mention several points which we have ignored but may affect the behavior of the noise and the autocorrelation. First, if there exist decoherence mechanism other than thermal agitation, it affects the decaying behavior. To investigate it experimentally is an important issue as it gives critical information on decoherence sources for the spin current. Second, the transmission may be energy-dependent in high temperature regime, while we assumed the energy-independent transmission here. We believe the assumption is valid in the energy scale of a few ten mK and a few μeV. For example, in the case of Fe/MgO, according to Fig.3 of Ref.[26], the typical energy scale governing the spin dependent transport is at least of the order of a few meV. So we believe that the energy dependence of the tunneling can be ignored if we consider such a very low energy scale. However, this assumption is certainly broken at high temperature (like room temperature) and the energy dependence becomes very important. Third, the transmission may be channel-dependent with some materials whose unique band structure can yield the channel dependence. For example, the conduction band of Fe/MgO has a strong spin dependence. In our calculation, this can be included as spin channel dependent transmission which is equivalent as including spin polarization of the electrodes as shown

before. We may experimentally detect the energy and channel dependence by checking the temperature dependence of Eq. (4) or changing materials, which is surely very interesting as it gives us new information. We may experimentally detect it through Eq.(4), which is surely very interesting as it conveys us new information.

In summary, according to the Landauer-Büttiker formalism, we have analytically derived the noise spectrum and the autocorrelation of the spin current at the tunnel junction. We show the temperature-independent behavior of the autocorrelation due to the Pauli exclusion principle for the spin current, which can be detectable experimentally. Such an experiment would enable us to directly address several unexplored aspects of the spin current, for example, its quantum coherence and dissipationless property.

**Acknowledgments**

This work was partially supported by JSPS KAKENHI Grant Numbers JP26220711, JP25103003, JP15H05854, JP16H05964, and JP26103002, Yazaki Memorial Foundation for Science and Technology, RIEC, Tohoku University.

**Captions**

**Figure 1** (a) Schematic of the lateral spin valve setup. Left electrode is ferromagnetic (FM), middle and right ones are nonmagnetic (NM1 and NM2). There is a tunnel barrier between NM1 and NM2. (b) Schematic of spin accumulation in the *x*-axis inside NM1. (c) Schematic of spin accumulation in the *z*-axis in the vicinity of the barrier between NM1 and NM2. We consider the noise generated at the barrier.

**Figure 2** Normalized noise spectrum $S_{\text{sym}}(\nu, T, \delta\mu)/S_0$ as a function of the normalized frequency $(h\nu/k_B T)$. Each curve shows noise spectrum with different spin accumulations $\delta\mu/k_B T = 0, 1, 3,$ and 5. We fix $\tau = 0.01$. All the curve converge to the zero temperature noise $S_{sym}(\nu, T = 0, \delta\mu)/S_0$ for $\nu \to \infty$. The dotted line shows the zero temperature spectrum of $S_{sym}(\nu, T = 0, \delta\mu = 0)/S_0$

**Figure 3** Autocorrelation of the spin current $C(t, \delta\mu, T)$ with various $h/k_B T$ ranging from 100 to 1000. $\delta\mu/k_B T = 5$ for solid curves, and $\delta\mu = 0$ for dotted lines. We fix $\tau = 0.01$. Clear oscillation with frequency $h/\delta\mu$ is observed.

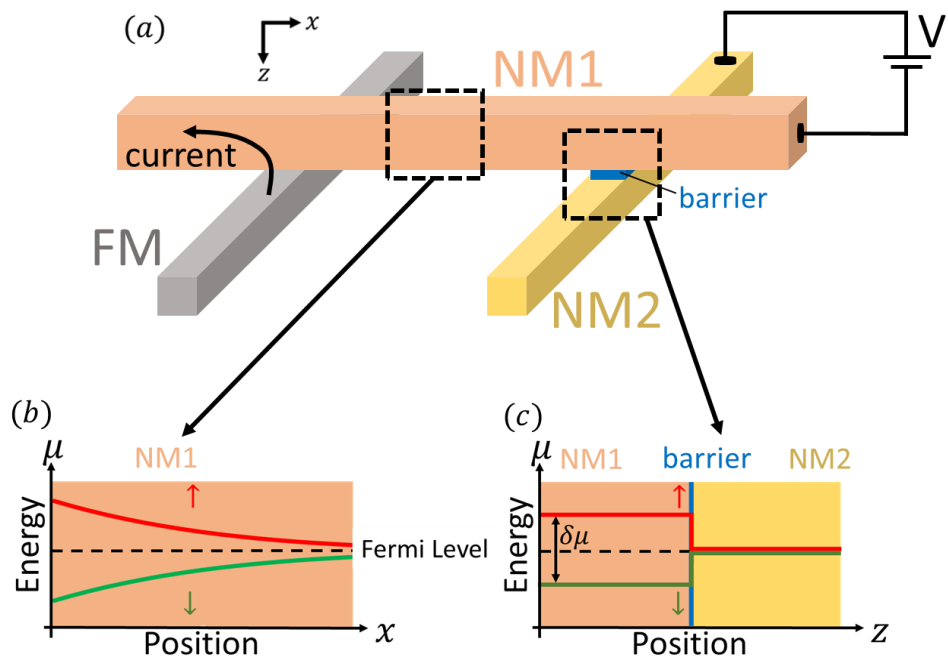

Figure 1 *Iwakiri et al.*

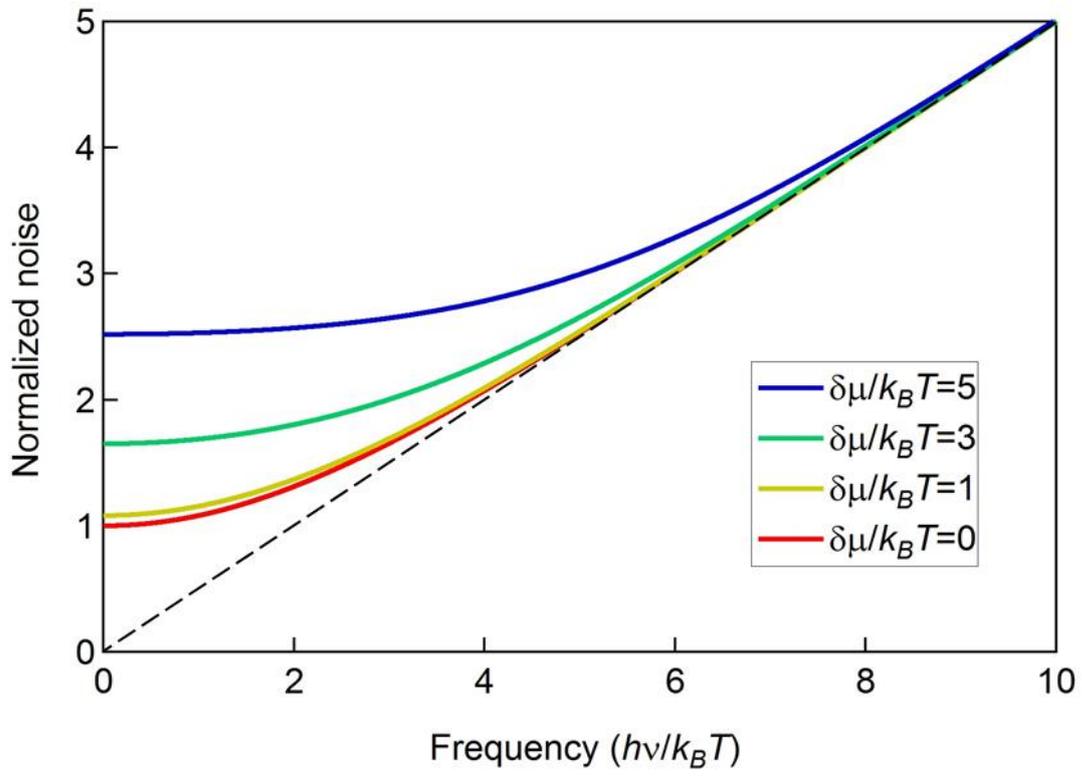

Figure 2 *Iwakiri et al.*

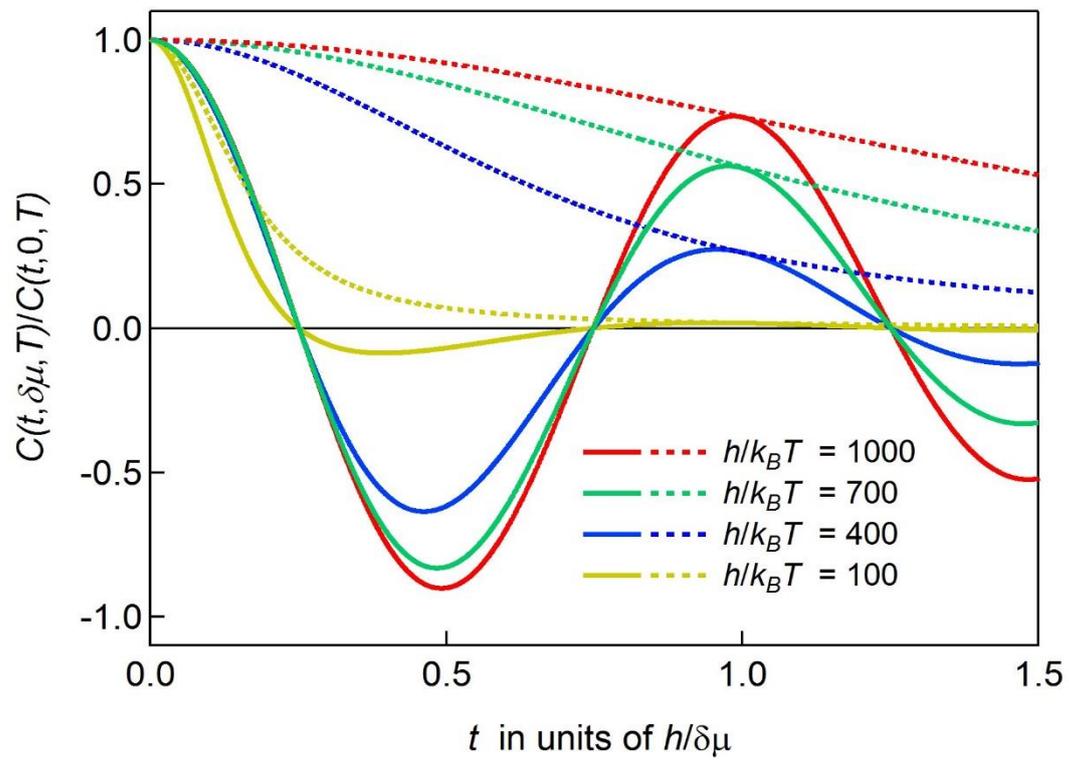

Figure 3 *Iwakiri et al.*